\documentclass[aps,12pt]{revtex4}

\newcommand{\be}{\begin{equation}}
\newcommand{\ee}{\end{equation}}
\usepackage{epsfig,amsmath,amssymb,graphics,graphicx}

\begin{document}

\begin{center}
{\it Physics of Plasmas 13 (2006) 052107 }
\end{center}

\title{Magnetohydrodynamics of Fractal Media}

\author{Vasily E. Tarasov }

\address{Skobeltsyn Institute of Nuclear Physics, 
Moscow State University, Moscow 119991, Russia } 

\email{E-mail: tarasov@theory.sinp.msu.ru}

\begin{abstract}
The fractal distribution of charged particles is considered.
An example of this distribution is 
the charged particles that are distributed over fractal.
The fractional integrals are used to describe fractal distribution.
These integrals are considered as approximations of
integrals on fractals.  
Typical turbulent media could be of a fractal structure
and the corresponding equations should be changed to include
the fractal features of the media.
The magnetohydrodynamics equations for fractal media are derived 
from the fractional generalization of integral Maxwell equations
and integral hydrodynamics (balance) equations. 
Possible equilibrium states for these equations are considered. 
\end{abstract}



\maketitle

\section{Introduction}

The theory of integrals and derivatives of noninteger order goes back 
to Leibniz, Liouville, Riemann, Grunwald, and Letnikov \cite{SKM,OS}. 
Fractional analysis has found many
applications in recent studies in mechanics and physics.
The interest in fractional integrals and derivatives 
has been growing continually 
during the last few years because of numerous applications. 
In a fairly short period of time the list of such 
applications becomes long, and include 
chaotic dynamics \cite{Zaslavsky1,Zaslavsky2},
material sciences \cite{Hilfer,C2,Nig1,Nig4}, 
mechanics of fractal and complex media 
\cite{Mainardi,Media,AP2005-2,Physica2005},
quantum mechanics \cite{Laskin,Naber}, 
physical kinetics \cite{Zaslavsky1,Zaslavsky7,SZ,ZE},
plasma physics \cite{CLZ,Plasma2005}, 
electromagnetic theory \cite{Lutzen,Mil2,Plasma2005},  
astrophysics \cite{CMDA},
long-range dissipation \cite{GM,TZ2}, 
non-Hamiltonian mechanics \cite{nonHam,FracHam},
long-range interaction \cite{Lask,TZ3}, 
anomalous diffusion, and transport theory 
\cite{Zaslavsky1,Montr,Uch,MK}.

The new type of problem has increased rapidly in areas 
in which the fractal features of a process or the medium 
impose the necessity of using nontraditional tools 
in "regular" smooth physical equations. 
To use fractional derivatives and fractional integrals 
for fractal distribution, we must use some 
continuous model \cite{Media,AP2005-2}. 
We propose to  describe the fractal medium by a fractional 
continuous model \cite{Media}, where all characteristics and 
fields are defined 
everywhere in the volume, but they follow some generalized 
equations that are derived by using fractional integrals.
In many problems the real fractal structure of the medium
can be disregarded and the fractal medium can be replaced by
some fractional continuous mathematical model.
Smoothing of microscopic characteristics over the
physically infinitesimal volume transforms the initial
fractal medium into the fractional continuous model \cite{Media,AP2005-2}
that uses the fractional integrals.
The order of the fractional integral is equal
to the fractal dimension of distribution. 
The fractional integrals allow us to take into account the
fractality of the distribution.
Fractional integrals can be considered as approximations 
of integrals on fractals \cite{Svozil,RLWQ}. 
In Ref. \cite{Svozil,RLWQ}, authors proved that integrals
on fractals can be approximated by fractional integrals.
In Ref. \cite{nonHam}, we proved that fractional integrals
can be considered as integrals over the space with a fractional
dimension up to the numerical factor. 

The distribution on the fractal can be described by 
a fractional continuous medium model
\cite{Media,AP2005-2,Physica2005,Plasma2005}. 
In the general case, the fractal medium cannot 
be considered as a continuous medium.
There are points and domains without particles.
In Refs. \cite{Media,AP2005-2,Plasma2005}, 
we suggest considering the fractal distributions  
as special (fractional) continuous media.
We use the procedure of replacement of the distribution 
with fractal mass dimension by some continuous model that 
uses fractional integrals. This procedure is a fractional 
generalization of the Christensen approach \cite{Chr}.
The suggested procedure leads to the fractional integration 
to describe the fractal medium.
In this paper, we consider the magnetohydrodynamics equations 
for the fractal distribution of charged particles.
Note that typical turbulent media could be of a fractal structure
and the corresponding equations should be changed to include
the fractal features of the media.

In Sec. II, a brief review of the Hausdorff measure, 
Hausdorff dimension and integration on fractals is suggested
to fix notation and provide a convenient reference.
The connection integration on fractals and fractional 
integration is discussed.
In Sec. III, a brief review of electrodynamics of 
the fractal distribution of charged particles is given. 
The densities of electric charge and current 
for the fractal distribution are described. 
A fractional generalization of the
integral Maxwell equation is suggested. 
In Sec. IV, a brief review of the hydrodynamics of 
fractal media is considered to 
fix notation and provide a convenient reference.
In Sec. V, the magnetohydrodynamics equations
for the fractal distribution of charged particles are derived.
The stationary states for these equations are considered.
Finally, a short conclusion is given in Sec. VI.

\section{Integration on fractal and fractional integration} 

Fractals are measurable metric sets 
with a noninteger Hausdorff dimension.  
The main property of the fractal is noninteger Hausdorff dimension.
Let us consider a brief review of 
the Hausdorff measure and the Hausdorff dimension \cite{Falconer,Feder} 
to fix the notation and provide a convenient reference. 

\subsection{Hausdorff measure and Hausdorff dimension}

Consider a measurable metric set $(W, \mu_H)$. 
The elements of $W$ are denoted by $x, y, z, . . . $, and represented by 
$n$-tuples of real numbers, $x = (x_1,x_2,...,x_n)$, 
such that $W$ is embedded in $\mathbb{R}^n$. 
The set $W$ is restricted by the following conditions:
(1) $W$ is closed;
(2) $W$ is unbounded; 
(3) $W$ is regular (homogeneous, uniform) with its points randomly distributed.

The metric $d(x,y)$ as a function of two points $x$ and $y \in W$ 
can be defined  by
\be
d(x,y)=\sum^n_{i=1} |y_i-x_i| ,
\ee
or
\be
d(x,y)=|x-y|=\left( \sum^n_{i=1}(y_i-x_i)^2 \right)^{1/2} .
\ee
The diameter of a subset $E \subset W \subset \mathbb{R}^n$ is 
\[
d(E)=diam(E)=\sup\{ d(x,y): \ x , y  \in E \}  ,
\]

Let us consider a set $\{E_i\}$ of non-empty 
subsets $E_i$ such that
$dim(E_i) < \varepsilon$,  $\forall i$, and 
$W \subset \bigcup^{\infty}_{i=1} E_i$.
Then, we define 
\be
\xi(E_i,D)= \omega(D) [diam (E_i)]^D=\omega(D)  [d(E_i)]^D .
\ee
The factor $\omega(D)$ 
depends on the geometry of $E_i$, used for covering $W$.
If $\{E_i\}$ is the set of all (closed or open) balls in $W$, then
\be
\omega(D)=\frac{\pi^{D/2} 2^{-D}}{\Gamma(D/2+1)}. 
\ee

The Hausdorff dimension $D$ of a subset $E \subset W$ 
is defined \cite{Federer,R,Ed,Falconer} by 
\be \label{Hd}
D= dim_H (E)=\sup \{ d \in R: \quad  \mu_H (E, d ) = \infty  \} ,
\ee
or
\be \label{Hd2}
D= dim_H (E)=\inf\{ d \in R : \quad \mu_H(E,d)=0 \}.
\ee
From (\ref{Hd}) and (\ref{Hd2}), we obtain 

\noindent
1) $\mu_H(E,d)=0$ for $d>D=dim_H(E)$;  \\
2) $\mu_H(E,d)=\infty$ for $d<D=dim_H(E)$. \\

The Hausdorff measure $\mu_H$ of a subset $E \subset W $ 
is defined \cite{Federer,R,Ed,Falconer} by
\be
\mu_H(E,D)=
\lim_{\varepsilon \rightarrow 0} \inf_{\{E_i\}} 
\{ \sum^{\infty}_{i=1} \xi(E_i,D): \quad 
E \subset \bigcup_i E_i , \quad d(E_i)< \varepsilon \quad \forall i \} ,
\ee
or
\be 
\mu_H(E,D)= \omega(D) \lim_{d(E_i) \rightarrow 0} 
\inf_{\{E_i\}} \sum^{\infty}_{i=1} [d(E_i)]^D .
\ee
If $E \subset W$ and $\lambda >0$, then
\[ \mu_H(\lambda E,D) =\lambda^D \mu_H(E,D) , \]
where $\lambda E =\{ \lambda x, \ x \in E \}$.

\subsection{Function and integrals on fractal}

Let us consider the functions on $W$: 
\be \label{f}
f(x)=\sum^{\infty}_{i=1} \beta_i \chi_{E_i}(x) , 
\ee
where $\chi_{E}$ is the characteristic function of $E$: 
$\chi_{E}(x)=1$ if $x \in E$, and $\chi_{E}(x)=0$ if $x \not \in E$. 
For continuous function $f(x)$:
\be
\lim_{x \rightarrow y} f(x)=f ( y ) 
\ee
whenever 
\be
\lim_{x \rightarrow y} d(x, y)=0 .
\ee

The Lebesgue-Stieltjes integral for (\ref{f}) 
is defined by
\be \label{LSI}
\int_W f d \mu =\sum^{\infty}_{i=1} \beta_i \mu_H(E_i).
\ee
Therefore
\[
\int_W f(x) d \mu_H (x) =
\lim_{ d(E_i) \rightarrow 0} \sum_{E_i} 
f(x_i) \xi(E_i,D)= 
\]
\be \label{int}
=\omega(D)  
\lim_{ d(E_i) \rightarrow 0} \sum_{E_i} f(x_i) [d(E_i)]^D .
\ee
It is always possible to divide $\mathbb{R}^n$ into parallelepipeds:  
\be
E_{i_1...i_n} =\{  (x_1,...,x_n) \in W: x_j =
(i_j-1) \Delta x_j +\alpha_j, \
0 \le \alpha_j \le \Delta x_j, \quad j=1,...,n \} .
\ee
Then
\[
d \mu_H (x)= \lim_{d(E_{i_1...i_n}) \rightarrow 0}
\xi(E_{i_1 ... i_n},D)=
\]
\be
=\lim_{d(E_{i_1 ... i_n}) \rightarrow 0}
\prod^n_{j=1} (\Delta x_j)^{D/n}=\prod^n_{j=1} d^{D/n} x_j .
\ee
The range of integration $W$ may also be parametrized by 
polar coordinates with $r=d(x, 0)$ and angle $\Omega$. 
Then $E_{r,\Omega}$ can be thought of as spherically 
symmetric covering around a center at the origin. 
In the limit, the function $\xi(E_{r,\Omega},D)$ gives
\be
d\mu_H(r,\Omega)=\lim_{d(E_{r, \Omega}) \rightarrow 0}
\xi (E_{r,\Omega},D)=d\Omega^{D-1} r^{D-1} dr. 
\ee

Let us consider $f(x)$ that is symmetric with respect 
to some center $x_0 \in W$, 
i.e., $f(x) = const$ for all $x$, such that $d(x, x_0)=r$ 
for arbitrary values of $r$. Then the transformation
\be \label{WrZ}
W \rightarrow W^{\prime} : \ x \rightarrow x^{\prime}=x-x_0
\ee
can be performed to shift the center of symmetry.  
Since $W$ is not a linear space, (\ref{WrZ}) need not be a map of $W$ onto 
itself; (\ref{WrZ}) is measure preserving. 
Then the  integral over a $D$-dimensional metric space is 
\be \label{intWf}
\int_W f d\mu_H = \lambda(D) \int^{\infty}_0 f(r) r^{D-1} dr ,
\ee
where
\be
\lambda(D)=\frac{2 \pi^{D/2}}{\Gamma(D/2)} .
\ee
This integral is known in the theory of the 
fractional calculus \cite{SKM}. 
The right Riemann-Liouville 
fractional integral is
\be \label{FID}
(I^{D}_{-} f)(z)=\frac{1}{\Gamma(D)} \int^{\infty}_z (x-z)^{D-1} f(x) dx .
\ee
For $z=0$, Eq. (\ref{FID}) gives
\be \label{ID-}
(I^{D}_{-} f)(0)=\frac{1}{\Gamma(D)} \int^{\infty}_0 x^{D-1} f(x) dx  ,
\ee
and Eq. (\ref{intWf}) is reproduced by
\be \label{FIFI}
\int_W f d\mu_H = \frac{2 \pi^{D/2} \Gamma(D)}{\Gamma(D/2)} (I^{D}_{-} f)(0) .
\ee
Equation (\ref{FIFI}) connects the integral on fractal 
with an integral of fractional order.
This result permits us to apply different tools of the fractional calculus
\cite{SKM} for the fractal medium.
As a result, the fractional integral can be considered as an
integral on the fractal up to the numerical factor 
$\Gamma(D/2) /[ 2 \pi^{D/2} \Gamma(D)]$.

Note that the interpretation of fractional integration
is connected with a fractional dimension \cite{nonHam}.
This interpretation follows from the well-known formulas 
for dimensional regularization \cite{Col}.
The fractional integral can be considered as an
integral in the fractional dimension space
up to the numerical factor $\Gamma(D/2) /[ 2 \pi^{D/2} \Gamma(D)]$.
In Ref. \cite{Svozil} it was proved that the fractal space-time 
approach is technically identical to the dimensional regularization.

\subsection{Properties of integrals}

The integral defined in (\ref{int}) satisfies the following properties:

(1) Linearity:
\be \label{14}
\int_W (af_1+bf_2) \, d \mu_H=a \int_W f_1 \, d \mu_H+b \int_W f_2 \, d \mu_H ,
\ee
where $f_1$ and $f_2$ are arbitrary functions; 
$a$ and $b$ are arbitrary constants.

(2) Translational invariance:
\be \label{15}
\int_W f(x+x_0) \, d \mu_H(x)= \int_W f(x) \, d \mu_H(x)
\ee
since $d \mu_H(x - x_0)=d \mu_H(x)$ as a consequence 
of homogeneity (uniformity).

(3) Scaling property:
\be \label{16}
\int_W f(\lambda x) \, d \mu_H(x)= \lambda^{-D}\int_W f(x) \, d \mu_H(x)
\ee
since $d \mu_H (x/\lambda)=\lambda^{-D}d \mu_H(x)$.

By evaluating the integral of the function 
$f(x) = \exp(-ax^2+bx)$, it has been shown \cite{Wilson,Col} 
that conditions (\ref{14})-(\ref{16}) define 
the integral up to normalization:
\be \label{18}
\int_W \exp(-ax^2+bx) d \mu_H (x)=
\pi^{D/2} a^{-D/2} \exp(b^2/4a) .
\ee
Note that, for $b = 0$, Eq. (\ref{18}) 
is identical to result from (\ref{FIFI}), 
which can be obtained directly without conditions (\ref{14})-(\ref{16}).

\subsection{Multi-variable integration on fractal}

The integral in (\ref{intWf}) is defined for a single variable.
It is only useful for integrating spherically symmetric functions. 
We consider multiple 
variables by using the product spaces and product measures.

Let us consider a collection of $n=3$ 
measurable sets $(W_k,\mu_k,D)$ with $k=1,2,3$,
and form a Cartesian product of the sets $W_k$ producing  
$W=W_1 \times W_2 \times W_3$.
The definition of product measures and the application of the Fubinis theorem 
provides a measure for the product set $W=W_1 \times W_2 \times W_3$ as
\be
(\mu_1 \times \mu_2 \times \mu_3)(W) 
= \mu_1(W_1) \mu_2(W_2) \mu_3(W_3). 
\ee
Then integration over a function $f$ on $W$ is
\[
\int_W f (x_1,x_2, x_3) d(\mu_1 \times \mu_2 \times \mu_3) = 
\]
\be \label{int-n}
=\int_{W_1} \int_{W_2} \int_{W_3} 
f (x_1,x_2 , x_3) d \mu_1(x_1) d \mu_2(x_2) d \mu_3(x_3). 
\ee
In this form, the single-variable measure from (\ref{intWf}) may 
be used for each coordinate $x_k$, which has 
an associated dimension $\alpha_k$:
\be 
d \mu_k(x_k) = \lambda(\alpha_k) |x_k|^{\alpha_k-1} dx_k , \quad k=1,2,3.
\ee
Then the total dimension of  $W=W_1 \times W_2 \times W_3$ is  
\be \label{Dk} D= \alpha_1+\alpha_2+\alpha_3  . \ee

Let us reproduce the result for the single-variable integration 
(\ref{intWf}), from $W_1  \times  W_2 \times W_3$.  
We take a spherically symmetric function $f (x_1, x_2 , x_3) = f (r)$,
where $r^2 = (x_1)^2 + (x_2)^2 + (x_3)^2$ and perform the integration 
in spherical coordinates $(r, \phi, \theta)$. 
Equation (\ref{int-n}) becomes
\[
\int_W d \mu_1(x_1) d\mu_2(x_2) d\mu_3(x_3) f (x_1,x_2,x_3) = 
\]
\[
= A(\alpha) \int_{W_1} d x_1 \int_{W_2} d x_2 \int_{W_3} 
d x_3 |x_1|^{\alpha_1-1} \,
|x_2|^{\alpha_2-1} |x_3|^{\alpha_3-1} f(x_1,x_2,x_3)= 
\]
\be
=A(\alpha) \int dr 
\int d \phi \int d \theta \, J_3 \, r^{ \alpha_1+\alpha_2+ \alpha_3- 3} 
(\cos \phi)^{\alpha_1-1} (\sin \phi)^{\alpha_2+ \alpha_3 -2} 
(\sin \theta)^{\alpha_3 -1}  f(r) ,
\ee
where $J_3= r^2 \sin \phi$ is the Jacobian of the coordinate change,
and $A(\alpha)=\lambda(\alpha_1) \lambda(\alpha_2) \lambda (\alpha_3)$.
Since the function is only dependent on the radial variable and 
not the angular variables, we can use 
\be
\int^{\pi/2}_0 \sin^{\mu-1} x \cos^{\nu-1} x dx =
\frac{\Gamma(\mu/2) \Gamma(\nu/2)}{2 \Gamma((\mu+\nu)/2)} .
\ee
where $\mu >0$, $\nu>0$.
From (\ref{Dk}), we obtain
\be
\int_W d\mu_1(x_1) d \mu_2(x_2) d\mu_3(x_3)f (r) = 
\lambda(D) \int f(r) r^{D-1} dr ,
\ee
where
\be
\lambda(D)=\frac{2 \pi^{D/2}}{\Gamma(D/2)} .
\ee
This equation describes the $D$-dimensional integration 
\cite{Col} of a spherically symmetric function, 
and reproduces the result (\ref{intWf}).

\subsection{Density function and mass on fractal}

Let us consider the mass that is distributed on the measurable metric set 
$W$ with the fractional Hausdorff dimension $D$.
Suppose that the density of mass distribution is described by the 
function $\rho({\bf r})$ that is defined by (\ref{f}).
In this case, the mass can be derived by 
\be
M_D(W)= \int_W \rho({\bf r}) dV_D, 
\ee 
where
\be
dV_D=d \mu_1(x_1) d \mu_2(x_2) d \mu_3(x_3)=c_3(D,{\bf r})dxdydz ,
\ee
\be
c_3(D,{\bf r})= \lambda(\alpha_1) \lambda(\alpha_2) \lambda(\alpha_3)
x^{\alpha_1-1} y^{\alpha_2-1}  z^{\alpha_3-1} ,
\ee
\be
dim_H (W)=D =\alpha_1+\alpha_2+\alpha_3 .
\ee
As a result, we have
\be \label{MD}
M_D(W)=\int_W \rho ({\bf r}) dV_D, \quad dV_D=c_3(D,{\bf r}) dV_3,
\ee 
where $dV_3=dx dy dz$ for Cartesian coordinates, and
\be
c_3(D,{\bf r})=\frac{8 \pi^{D/2}
|x|^{\alpha_1-1} |y|^{\alpha_2-1} |z|^{\alpha_3-1} }{\Gamma(\alpha_1) 
\Gamma(\alpha_2) \Gamma(\alpha_3) } .
\ee
As a result, we get the Riemann-Liouville fractional integral \cite{SKM}
up to a numerical factor $8 \pi^{D/2}$.


\subsection{Mass of fractal distribution}

The cornerstone of fractals is the noninteger dimension. 
The fractal dimension can be best calculated by box counting method, 
which means drawing a box of size $R$ 
and counting the mass inside. 
This mass fractal dimension can be easy measured
for fractal media.
The properties of the fractal medium like mass obey
a power law relation $M \sim R^{D}$, 
where $M$ is the mass of the fractal medium, $R$ 
is a box size (or a sphere radius),
and $D$ is a mass fractal dimension. 
The power law relation $M \sim R^{D}$ can be naturally 
derived by using the fractional integral \cite{Media}.
The mass fractal dimension is connected \cite{Media} 
with the order of fractional integrals.

Consider the region $W$ in three-dimensional 
Euclidean space $\mathbb{R}^3$.
The volume of the region $W$ is denoted by $V_D(W)$.
The mass of the region $W$ in the fractal medium is denoted by $M_D(W)$. 
The fractality of the medium means than the mass of this medium 
in any region $W$ of Euclidean space $\mathbb{R}^3$ increases 
more slowly than the volume of this region.
For the ball region of the fractal medium, 
this property can be described by the power law $M \sim R^{D}$, 
where $R$ is the radius of the ball $W$. 

The fractal medium is called a homogeneous one if the power 
law $M \sim R^{D}$ does not depend on the translation  
of the region. The homogeneity property of the medium
can be formulated in the form:
For all regions $W$ and $W^{\prime}$ of the homogeneous fractal medium 
with the equal volumes $V_D(W)=V_D(W^{\prime})$, 
the masses of these regions are equal $M_D(W)=M_D(W^{\prime})$. 
Note that the wide class of the fractal media satisfies 
the homogeneous property.

In Refs. \cite{Media}, the continuous medium model for the fractal media was
suggested. The fractality and homogeneity properties can be 
realized in the following forms: 
(1) Homogeneity:
The local density of homogeneous fractal media
is a translation invariant value that has the form
$\rho({\bf r})=\rho_0=const$.
(2) Fractality:
The mass of the ball region $W$ of a fractal medium obeys a power law relation,
$M \sim R^{D}$, where $0<D<3$, and $R$ is the radius of the ball.
These requirements can be realized by the 
fractional generalization (\ref{MD}) of the equation
\be \label{MW} M_3(W)=\int_W \rho({\bf r}) d V_3  . \ee
The form of function $c_3(D,{\bf r})$ is defined by the 
properties of the fractal medium.   
Note that the final equations that relate the physical variables 
have a form that is independent of a numerical factor 
in the function $c_3(D,{\bf r})$. 
However, the dependence of ${\bf r}$ is important to these equations. 

Equation (\ref{MD}) describes the mass that is distributed in the volume
and has the mass fractal dimension $D$ by fractional integrals.
There are many different definitions of fractional integrals \cite{SKM}.
The fractional integrals can be used to describe fields that 
are defined on the set $W$ with fractional Hausdorff dimension $dim_H(W)=D$.

For the Riemann-Liouville fractional integral, 
\be \label{c3Dr}
c_3(D,{\bf r})=\frac{ |x|^{\alpha_1-1}|y|^{\alpha_2-1} |z|^{\alpha_3-1} }{
\Gamma(\alpha_1) \Gamma(\alpha_2) \Gamma(\alpha_3) },
\ee
where $x$, $y$, $z$ are Cartesian's coordinates, 
and $D=\alpha_1+\alpha_2+\alpha_3$, $0<D\le 3$.

Note that for $D=2$, we have the fractal mass distribution in the volume. 
In general, this case is not equivalent to the distribution
on the two-dimensional surface.

For $\rho({\bf r})=\rho(|{\bf r}|)$ , we can use
the Riesz definition of the fractional integrals \cite{SKM}, and
\be \label{IDc}
c_3(D,{\bf r})=\lambda(D) |{\bf r}|^{D-3} , \ee
where
\be
\lambda(D) =\gamma^{-1}_3(D)=\frac{\Gamma(1/2)}{2^D \pi^{3/2} \Gamma(D/2)} . 
\ee
Note that
\be
\lim_{D\rightarrow 3-} \gamma^{-1}_3(D) = (4 \pi^{3/2})^{-1} .
\ee
Therefore, we suggest using
\be \label{46}
\lambda(D) =(4 \pi^{3/2})  \gamma^{-1}_3(D)= 
\frac{2^{3-D} \Gamma(3/2)}{\Gamma(D/2)} .
\ee
The factor (\ref{46}) allows us to derive the usual integral 
in the limit $D\rightarrow (3-0)$.
Note that the final equations that relate mass, moment of inertia, 
and radius are independent of the numerical factor $\lambda(D)$. 

For the homogeneous medium ($\rho({\bf r})=\rho_0=const$) and 
the ball region $W=\{{\bf r}: \  |{\bf r}|\le R \}$,  
\[ M_D(W)= \rho_0 \frac{2^{3-D} \Gamma(3/2)}{\Gamma(D/2)} 
\int_W |{\bf r}|^{D-3} d V_3 . \]
Using the spherical coordinates, we get
\[ M_D(W)= \frac{\pi 2^{5-D} \Gamma(3/2)}{\Gamma(D/2)} \rho_0 
\int_W |{\bf r}|^{D-1} d |{\bf r}|= 
\]
\[ =\frac{2^{5-D} \pi \Gamma(3/2)}{D \Gamma(D/2)} \rho_0 R^{D} . \]
As a result, we have $M(W)\sim R^D$, i.e., we derive the equation 
$M \sim R^{D}$ up to the numerical factor.
Therefore the fractal medium with noninteger mass dimension $D$ can be
described by fractional integral of order $D$.

\section{Electrodynamics of fractal distribution of charged particles}

In this section, a brief review of electrodynamics of 
fractal distribution of charged particles \cite{Plasma2005}
is considered to fix notation and provide a convenient reference.

\subsection{Electric charge for fractal distribution}

Let us consider charged particles that are
distributed with a constant density over a fractal with
Hausdorff dimension $D$. 
In this case, the electric charge $Q$ satisfies the 
scaling law $Q(R) \sim R^{D}$,
whereas for a regular n-dimensional Euclidean object 
we have $Q(R)\sim R^n$. 

The total charge of region $W$ is 
\be\label{QW}
Q_3(W)=\int_W \rho({\bf r},t) dV_3 , \ee
where $\rho({\bf r},t)$ is a charge density in the region $W$. 
The fractional generalization of (\ref{QW}) is
\[ Q_D(W)=\int_W \rho({\bf r},t) dV_D , \]
where $D$ is a fractal dimension of the distribution, and 
\be \label{5a} dV_D=c_3(D,{\bf r})dV_3. \ee
The functions $c_3(D,{\bf r})$ is defined by the properties
of the distribution. 

If we consider the ball region $W=\{{\bf r}: \ |{\bf r}|\le R \}$, 
and spherically symmetric distribution of charged particles 
($\rho({\bf r},t)=\rho(r,t)$), then 
\[ Q_D(R)=4\pi \frac{2^{3-D}\Gamma(3/2)}{\Gamma(D/2)}
\int^R_0 \rho(r) r^{D-1} dr . \]
For the homogeneous case, $\rho(r,t)=\rho_0$, and 
\[ Q_D(R)=4\pi \rho_0 \frac{2^{3-D}\Gamma(3/2)}{\Gamma(D/2)}
\frac{R^D}{D} \sim R^D . \]
The distribution of charged particles is called a homogeneous one
if all regions $W$ and $W^{\prime}$ with the equal 
volumes $V_D(W)=V_D(W^{\prime})$ have 
the equal total charges on these regions, $Q_D(W)=Q_D(W^{\prime})$.

\subsection{Electric current of fractal distribution}

For charged particles with density $\rho({\bf r},t)$ flowing 
with velocity ${\bf u}={\bf u}({\bf r},t)$, 
the current density ${\bf J}({\bf r},t)$ is 
\[ {\bf J}({\bf r},t)= \rho({\bf r},t) {\bf u} . \]
The electric current $I(S)$ is defined as the flux of electric charge.
Measuring the field ${\bf J}({\bf r},t)$ passing through a surface 
$S=\partial W$ gives  
\be \label{IS}
I(S)=\Phi_J(S)=\int_S ({\bf J}, d{\bf S}_2) , \ee
where $d{\bf S_2}=dS_2{\bf n}$ is a differential unit of area 
pointing perpendicular to the surface $S$, 
and the vector ${\bf n}=n_k {\bf e}_k$ is a vector of normal.
The fractional generalization of (\ref{IS}) is
\[ I(S)=\int_S ({\bf J}({\bf r},t), d{\bf S}_d) , \]
where 
\be \label{C2} dS_d=c_2 (d,{\bf r})dS_2 , \quad 
c_2(d,{\bf r})= \frac{2^{2-d}}{\Gamma(d/2)} |{\bf r}|^{d-2} . \ee
Note that $c_2(2,{\bf r})=1$ for $d=2$. 
The boundary $\partial W$ has the dimension $d$. 
In general, the dimension $d$ is not equal to $2$ and 
is not equal to $(D-1)$.

\subsection{Charge conservation for fractal distribution}

The electric charge has a fundamental property established
by numerous experiments: the velocity of charge change
in region $W$ bounded by the surface $S=\partial W$
is equal to the flux of charge through this surface.
This is known as the law  of charge conservation:
\[ \frac{dQ(W)}{dt}=-I(S), \]
or, in the form
\be \label{cecl} \frac{d}{dt} \int_W \rho({\bf r},t) dV_D= 
- \oint_{\partial W} ({\bf J} ({\bf r},t),d{\bf S}_d) . \ee
In particular, when the surface $S=\partial W$ is fixed,
we can write
\be \label{drho} \frac{d}{dt} \int_W \rho({\bf r},t) dV_D= 
\int_W \frac{\partial \rho({\bf r},t)}{\partial t} dV_D .\ee
Using the fractional generalization of the 
Gauss's theorem (see the Appendix), we get
\[ 
\oint_{\partial W} ({\bf J} ({\bf r},t),d{\bf S}_d) =
\]
\be \label{gt}
=\int_W c^{-1}_3(D,{\bf r})
\frac{\partial}{\partial x_k} \Bigl( c_2(d,{\bf r})J_k({\bf r},t) \Bigr)
dV_D .\ee
The substitution of Eqs. (\ref{drho}) and (\ref{gt})
into Eq. (\ref{cecl}) gives 
\be \label{53}
c_3(D,{\bf r})\frac{\partial \rho({\bf r},t)}{\partial t}+
\frac{\partial}{\partial x_k} \Bigl( c_2(d,{\bf r})J_k({\bf r},t) \Bigr)=0. 
\ee
As a result, we obtain the law of charge 
conservation in differential form (\ref{53}).
This equation can be considered as a continuity equation for
fractal distribution of particles \cite{AP2005-2}.

\subsection{Electric field and Coulomb's law}

For a point charge $Q$ at position  ${\bf r}^{\prime}$, 
the electric field at a point ${\bf r}$ is defined by
\[ {\bf E}=\frac{Q}{4 \pi \varepsilon_0} \
\frac{{\bf r}-{\bf r}^{\prime}}{|{\bf r}-{\bf r}^{\prime}|^3} , \]
where $\varepsilon_0$
is a fundamental constant called the permittivity of free space. 

For a continuous stationary distribution $\rho({\bf r}^{\prime})$,   
\be \label{E}
{\bf E}({\bf r})=\frac{1}{4 \pi \varepsilon_0} \int_W
\frac{{\bf r}-{\bf r}^{\prime}}{|{\bf r}-{\bf r}^{\prime}|^3}
\rho({\bf r}^{\prime}) dV^{\prime}_3 .
\ee
For Cartesian's coordinates 
$dV^{\prime}_3=dx^{\prime}dy^{\prime}dz^{\prime}$.
The fractional generalization of (\ref{E}) is
\be \label{CLD}
{\bf E}({\bf r})=\frac{1}{4 \pi \varepsilon_0} \int_W
\frac{{\bf r}-{\bf r}^{\prime}}{|{\bf r}-{\bf r}^{\prime}|^3}
\rho({\bf r}^{\prime}) dV^{\prime}_D , \ee
where $dV^{\prime}_D=c_3(D,{\bf r}^{\prime}) dV^{\prime}_3$. 
Equation (\ref{CLD}) can be considered as Coulomb's law 
for a fractal stationary distribution of electric charges. 

The electric field passing through a surface 
$S=\partial W$ gives the electric flux 
\[ \Phi_E(S)=\int_S ({\bf E}, d{\bf S}_2) , \]
where ${\bf E}$ is the electric field vector, and $d{\bf S}_2$ 
is a differential unit of area pointing perpendicular to the surface S.

\subsection{Gauss's law for fractal distribution}

Gauss's law tells us that the total flux $\Phi_E(S)$ of 
the electric field ${\bf E}$
through a closed surface $S=\partial W$ 
is proportional to the total electric charge $Q(W)$
inside the surface: 
\be \label{GL1} \Phi_E(\partial W)=\frac{1}{\varepsilon_0} Q(W) . \ee
For the fractal distribution, Gauss's law (\ref{GL1}) states 
\be \label{GL2} \int_S ({\bf E},d{\bf S}_2)=\frac{1}{\varepsilon_0} 
\int_W \rho ({\bf r},t) dV_D , \ee
where ${\bf E}={\bf E}({\bf r},t)$ is the electric field, and 
$\rho({\bf r},t)$ is the 
charge density,  $dV_D=c_3(D,{\bf r})dV_3$,
and $\varepsilon_0$ is the permittivity of free space.

If $\rho({\bf r},t)=\rho(r)$, and $W=\{{\bf r}:\ |{\bf r}|\le R\}$, then 
\[ Q(W)=4 \pi \int^R_0 \rho(r) c_3(D,{\bf r}) r^2 dr , \]
where 
\be \label{5R} c_3(D,{\bf r})=
\frac{2^{3-D}\Gamma(3/2)}{\Gamma(D/2)} |{\bf r}|^{D-3} . \ee
Then
\be \label{QW2} Q(W)=4 \pi \frac{2^{3-D}\Gamma(3/2)}{\Gamma(D/2)}
\int^R_0 \rho(r) r^{D-1} dr . \ee
For the sphere $S=\partial W=\{{\bf r}: \ |{\bf r}|= R \}$,
\be \label{PW} \Phi_E(\partial W)= 4 \pi R^2 E(R). \ee
Substituting (\ref{QW2}) and (\ref{PW}) in (\ref{GL1}),
we get 
\[ E(R)=\frac{2^{3-D}\Gamma(3/2)}{\varepsilon_0 R^2 \Gamma(D/2)}
\int^R_0 \rho(r) r^{D-1} dr .\]
For homogeneous ($\rho(\bf r)=\rho$) distribution, 
\[ E(R)=\rho \frac{2^{3-D}\Gamma(3/2)}{\varepsilon_0 D \Gamma(D/2)} 
R^{D-2}  \sim R^{D-2} .\]

\subsection{Magnetic field and Biot-Savart law}
    
The Biot-Savart law relates magnetic fields to the currents 
that are their sources. 
For a continuous distribution, the law is
\be \label{BSL0} {\bf B}({\bf r})=\frac{\mu_0}{4\pi} \int_W 
\frac{[{\bf J}({\bf r}^{\prime}),{\bf r}-{\bf r}^{\prime}]}{
|{\bf r}-{\bf r}^{\prime}|^3} d V^{\prime}_3 , \ee
where $[\ , \ ]$ is a vector product, 
${\bf J}$ is the current density, and 
$\mu_0$ is the permeability of free space. 
The fractional generalization of Eq. (\ref{BSL0}) is
\be \label{BSL} {\bf B}({\bf r})=\frac{\mu_0}{4\pi} \int_W 
\frac{[{\bf J}({\bf r}^{\prime}),{\bf r}-{\bf r}^{\prime}]}{
|{\bf r}-{\bf r}^{\prime}|^3} d V^{\prime}_D . \ee
This equation is the Biot-Savart law written 
for a steady current with a fractal distribution of electric charges.
The law (\ref{BSL}) can be used to find the magnetic 
field produced by any fractal distribution of steady currents.

\subsection{Ampere's law for fractal distribution}

The magnetic field in space around an electric current 
is proportional to the electric current that serves as its source. 
In the case of a static electric field, 
the line integral of the magnetic field around 
a closed loop is proportional to the electric current 
flowing through the loop. Ampere's law 
is equivalent to the steady state of the integral Maxwell equation 
in free space, and relates the spatially varying magnetic field 
${\bf B}({\bf r})$ to the current density ${\bf J}({\bf r})$. 

Note that, as mentioned in Ref. \cite{Lutzen}, 
Liouville, who was one of the pioneers in the development of 
fractional  calculus, was inspired by the problem of 
fundamental force law in Ampbre's electrodynamics and 
used fractional differential equations in that problem.  

Ampere's law states that
the line integral of the magnetic field ${\bf B}$ along 
the closed path $L$ around a current given in MKS by
\[ \oint_L ({\bf B},d{\bf l})=\mu_0 I(S) , \]
where $d{\bf l}$ is the differential length element.  
For the distribution of particles on the fractal, 
\[ I(S)=\int_S ({\bf J},d{\bf S}_d) , \]
where $d {\bf S}_d=c_2(d,{\bf r}) d S_2$.
For the cylindrically symmetric distribution, 
\[ I(S)=2 \pi \int^R_0 J(r) c_2(d,{\bf r}) r dr , \]
where $c_2(d,{\bf r})$ is defined in Eq. (\ref{C2}), i.e., 
\[ I(S)=4 \pi \frac{2^{2-d}}{\Gamma(d/2)}
\int^R_0 J(r) r^{d-1} dr . \]
For the circle $L=\partial W=\{{\bf r}: \ |{\bf r}|=R \}$, we get
\[ \oint_L ({\bf B},d{\bf l})= 2 \pi R \ B(R). \]
As a result, 
\[ B(R)= \frac{ \mu_0 2^{2-d}}{R \Gamma(d/2)}
\int^R_0 J(r) r^{d-1} dr .\]
For the homogeneous distribution, $J(r)=J_0$, and
\[ B(R)=J_0 \frac{\mu_0 2^{2-d}}{d \Gamma(d/2)} R^{d-1} \sim R^{d-1} .\]

\subsection{Fractional integral Maxwell equations} 

Let us consider the fractional integral 
Maxwell equations \cite{Plasma2005}.
The Maxwell equations are the set of 
fundamental equations for electric and magnetic fields.
The equations that can be expressed in integral form
are known as Gauss's law, Faraday's law, 
the absence of magnetic monopoles, and Ampere's law 
with displacement current.
In MKS,  these become 
\[ \oint_S ({\bf E},d{\bf S}_2)=
\frac{1}{\varepsilon_0} \int_W \rho dV_D , 
\]
\[ \oint_L ({\bf E},d{\bf l}_1)=
-\frac{\partial}{\partial t} \int_S ({\bf B},d{\bf S}_2) , \]
\[ \oint_S ({\bf B},d{\bf S}_2)= 0, \]
\[
\oint_L ({\bf B},d{\bf l}_1)=\mu_0 \int_S ({\bf J}, d{\bf S}_d)
+ \varepsilon_0 \mu_0\frac{\partial}{\partial t}
\int_S ({\bf E},d{\bf S}_2) . \]
Let us consider the fields that are defined on the fractal \cite{Feder} only.
The hydrodynamic and thermodynamics fields 
can be defined in the fractal media \cite{AP2005-2,Physica2005}.
Suppose that the electromagnetic field is defined on the fractal 
as an approximation of some real case with a fractal medium.
If the electric and magnetic fields are defined on a fractal and 
does not exist outside of the fractal in Eucledian space $\mathbb{R}^3$, 
then we must use the fractional generalization of the
integral Maxwell equations in the form \cite{Plasma2005}:
\[ \oint_S ({\bf E},d{\bf S}_d)=
\frac{1}{\varepsilon_0} \int_W \rho dV_D ,
\]
\[ 
\oint_L ({\bf E},d{\bf l}_{\gamma})=
-\frac{\partial}{\partial t} \int_S ({\bf B},d{\bf S}_d) , \]
\[
\oint_S ({\bf B},d{\bf S}_d)= 0, 
\]
\be \label{Max1}
\oint_L ({\bf B},d{\bf l_{\gamma}})=\mu_0 \int_S ({\bf J}, d{\bf S}_d)
+ \varepsilon_0 \mu_0\frac{\partial}{\partial t}
\int_S ({\bf E},d{\bf S}_d) . \ee
Note that fractional integrals are considered as an approximation 
of integrals on fractals \cite{Svozil,RLWQ}. 

Using the fractional generalization of Stokes's and Gauss's
theorems (see the Appendix), 
we can rewrite Eqs. (\ref{Max1}) in the form
\[ \int_W c^{-1}_3(D,{\bf r}) div( c_2(d,{\bf r}) {\bf E}) dV_D 
= \frac{1}{\varepsilon_0} \int_W \rho dV_D , \]
\[ \int_S  c^{-1}_2(d,{\bf r})
( curl(c_1(\gamma,{\bf r}){\bf E}), d{\bf S}_d) = 
-\frac{\partial}{\partial t} \int_S ({\bf B},d{\bf S}_d) , \]
\[ \int_W c^{-1}_3(D,{\bf r}) div(c_2(d,{\bf r}) {\bf B}) dV_d=0, \]
\[ \int_S c^{-1}_2(d,{\bf r}) 
(curl(c_1(\gamma,{\bf r}){\bf B}), d{\bf S}_d) = 
\mu_0 \int_S ({\bf J}, d{\bf S}_d)+ 
\varepsilon_0 \mu_0\frac{\partial}{\partial t}
\int_S ({\bf E},d{\bf S}_d) .  \]
As a result, we obtain  
\[ div \Bigl(c_2(d,{\bf r}) {\bf E} \Bigr) = 
\frac{1}{\varepsilon_0} c_3(D,{\bf r}) \rho  , \]
\[ curl \Bigl(c_1(\gamma,{\bf r}){\bf E} \Bigr)= 
-c_2(d,{\bf r}) \frac{\partial}{\partial t} {\bf B} , \]
\[ div \Bigl( c_2(d,{\bf r}) {\bf B} \Bigr)= 0, \]
\[  curl \Bigl( c_1(\gamma,{\bf r}){\bf B} \Bigr) = 
\mu_0 c_2(d,{\bf r}) {\bf J}+\varepsilon_0 \mu_0 c_2(d,{\bf r})
\frac{\partial {\bf E}}{\partial t}. \]

Note that the law of absence of magnetic monopoles 
for the fractal leads us to the equation 
$div ( c_2(d,{\bf r}) {\bf B} )= 0 $. 
It can be rewritten as
\[ div{\bf B}=-({\bf B}, grad c_2(d,{\bf r}) ) . \]
In general ($d \not=2$), the vector
$grad \ (c_2(d,{\bf r}))$ is not equal 
to zero and the magnetic field satisfies $div {\bf B}\not=0$.
If $d=2$, we have $div ({\bf B})\not=0$ only for nonsolenoidal 
field ${\bf B}$. Therefore the magnetic field on the fractal  
is similar to the nonsolenoidal field. 
As a result, the magnetic field on the fractal can be considered as 
a field with some "fractional magnetic monopole", 
$q_m\sim ({\bf B},\nabla c_2)$.


\section{Hydrodynamics of fractal media.}

\subsection{Euler equations for fractal media}

In Ref. \cite{AP2005-2}, we derive the fractional generalizations 
of integral balance equations for fractal media.
These equations leads to the following differential equations.  \\
(1) The equation of continuity, 
\be \label{1eq} \Bigl(\frac{d}{dt}\Bigr)_D \rho=-\rho \nabla^D_k u_k . \ee
(2) The equation of balance of density of momentum,  
\be \label{2eq} \rho \Bigl(\frac{d}{dt}\Bigr)_D u_k=
\rho f_k+\nabla^D_l p_{kl} . \ee
(3) The equation of balance of density of energy, 
\be \label{3eq} \rho\Bigl(\frac{d}{dt}\Bigr)_D e= c(D,d,R)
p_{kl} \frac{\partial u_k}{\partial x_l} . \ee
Here, we mean the sum on the repeated index, $k$ and $l$ from 1 to 3, 
and use the notations
\be \label{nabla}
\nabla^D_k A= a(D,d) R^{3-D} \frac{\partial }{\partial x_k}  
\Bigl( R^{d-2} A \Bigr) . \ee
\[ 
\Bigl(\frac{d}{dt}\Bigr)_D=\frac{\partial}{\partial t}+
c(D,d,R) u_l \frac{\partial}{\partial x_l} =
\]
\be
=\frac{\partial}{\partial t}+
a(D,d) R^{d+1-D} u_l \frac{\partial}{\partial x_l} . \ee
where 
\be \label{c} 
c(D,d,R)=a(D,d) R^{d+1-D}, 
\ee 
\be \label{a}
a(D,d)=\frac{2^{D-d-1} \Gamma(D/2)}{\Gamma(3/2) \Gamma(d/2)} .
\ee
The equations of balance are a set of five equations,
which are not closed.
These equations, in addition to the hydrodynamic fields
$\rho({\bf R},t)$, $u({\bf R},t)$, $e({\bf R},t)$, include
also the tensor of viscous stress $p_{kl}({\bf R},t)$.
Let us consider the special cases of (\ref{1eq})-(\ref{3eq}) with
\[ p_{kl}=- p \delta_{kl} , \]
where $p=p({\bf R},t)$ is the pressure. 
Then the hydrodynamics equations (\ref{1eq})-(\ref{3eq}) are
\be \label{Ee1} \Bigl(\frac{d}{dt}\Bigr)_D \rho=-\rho \nabla^D_k u_k . \ee
\be \label{Ee2} \Bigl(\frac{d}{dt}\Bigr)_D u_k=
f_k- \frac{1}{\rho}\nabla^D_k p . \ee
\be \label{Ee3} \Bigl(\frac{d}{dt}\Bigr)_D e= - c(D,d,R) \frac{p}{\rho}
\frac{\partial u_k}{\partial x_k}  . \ee
These equations are the Euler equations for the fractal medium.

\subsection{Equilibrium equation for fractal distribution}

The equilibrium state of medium means that 
\[ \frac{\partial A}{\partial t}=0, \quad 
\frac{\partial A}{\partial x_k}=0 , \]
for $A=\{\rho, u_k, e \}$.
In this case, Eqs. (\ref{Ee1})-(\ref{Ee3}) give
\be \label{ee1}
f_k=\frac{1}{\rho}\nabla^D_k p. 
\ee
Equation (\ref{ee1}) gives the fractional generalization of the 
equilibrium equations.
From (\ref{nabla}), (\ref{c}) and (\ref{a}), Eq. (\ref{ee1}) is
\[ \frac{\partial (c_2(d,R) p)}{\partial x_k}=\rho c_3(D,R) f_k . \]
For the homogeneous medium $\rho(x)=const$, and
\[ c_3(D,R)f_k=\frac{\partial (c_2(d,R)p/\rho_0) }{\partial x_k} . \]
If $c_3(D,R)f_k=-\partial U/ \partial x_k$, 
then 
\be \label{eqeq1} c_2(d,R)p+\rho_0 U =const. \ee
This equation is a fractional generalization of the equilibrium equation.

\subsection{Fractional Bernoulli integral}

Let us consider Eq. (\ref{Ee2}). 
Using Eq. (\ref{2eq}) and 
\[ \Bigl(\frac{d}{dt}\Bigr)_D \frac{{\bf u}^2}{2}=
u_k \Bigl(\frac{d}{dt}\Bigr)_D u_k , \]
we get 
\be \label{Bi1} \Bigl(\frac{d}{dt}\Bigr)_D \frac{{\bf u}^2}{2}=
u_k f_k-\frac{1}{\rho} u_k \nabla^D_k p . \ee
If 
\[ {\partial U}/{\partial t}=0 , \quad  
{\partial p}/{\partial t}=0, \]
then 
\be \label{d/dt} \Bigl(\frac{d}{dt}\Bigr)_D = c(D,d,R) \frac{d}{dt} . \ee
Suppose 
\be \label{npf} f_k=-c(D,d,R)\partial U/ \partial x_k . \ee 
If $D=3$ and $d=2$, then this force is potential. 
Using Eqs. (\ref{d/dt}) and (\ref{npf}), we get 
Eq. (\ref{Bi1}) in the form
\[ \frac{d}{dt}\Bigl( \frac{{\bf u}^2}{2}+U+P(d) \Bigr)=0 , \]
where 
\[ P(d)=\int^p_{p_0} \frac{d(c_2(d,R)p)}{c_2(d,R)\rho} .  \]
As a result, we obtain 
\be \label{IntB}
\sum^3_{k=1}\frac{u^2_k}{2}+U+P(d) =const . \ee
This integral of motion can be considered as a fractional generalization
of the Bernoulli integral for fractal media. 
If the forces $f_k$ are potential, and $D\not=3$, 
then the fractional analog 
of the Bernoulli integral does not exist.

For the density 
\be \label{rhoc2}
\rho=\rho_0 c^{-1}_2(d,R)=\rho_0 \frac{\Gamma(d/2)}{2^{2-d}} R^{2-d} , \ee
the integral (\ref{IntB}) gives
\be \label{Eq-i}
\frac{\rho_0 {\bf u}^2}{2}+\rho_0 U+ c_2(d,R) p = const . \ee
For $u_k=0$, Eq. (\ref{Eq-i}) leads to Eq. (\ref{eqeq1}).

\subsection{Sound waves in fractal media}

Let us consider the small perturbations 
of Eqs. (\ref{Ee1}) and (\ref{Ee2}):
\be \label{sp} \rho=\rho_0+\rho^{\prime}, \quad p=p_0+p^{\prime}, \quad 
u_k=u^{\prime}_k, \ee
where $\rho^{\prime} \ll \rho_0$, and $p^{\prime} \ll p_0$, and 
$p_0$ and $\rho_0$ describe the steady state:
\[ \frac{\partial \rho_0}{\partial t}=0,\quad
\frac{\partial \rho_0}{\partial x_k}=0, \quad
\frac{\partial p_0}{\partial t}=0, \quad
\frac{\partial p_0}{\partial x_k}=0 . \]
Supposing $f_k=0$, and  
substituting (\ref{sp}) into Eqs. (\ref{Ee1}) and (\ref{Ee2}), we get
\be \label{em3} \frac{\partial \rho^{\prime}}{\partial t}
=-\rho_0 \nabla^D_k u^{\prime}_k , \ee
\be \label{em4} \frac{\partial u^{\prime}_k}{\partial t}=
- \frac{1}{\rho_0}\nabla^D_k p^{\prime} . \ee
To derive the independent equations for $\rho^{\prime}$,  
we consider the partial derivative of Eq. (\ref{em3})
with respect to time:
\be \label{em5} \frac{\partial^2 \rho^{\prime}}{\partial t^2}
=-\rho \nabla^D_k \frac{ \partial u^{\prime}_k}{\partial t} . \ee
The substitution of (\ref{em4}) into (\ref{em5}) obtains
\be \label{em6} \frac{\partial^2 \rho^{\prime}}{\partial t^2}
=\nabla^D_k \nabla^D_k p^{\prime} . \ee
For adiabatic processes $p=p(\rho,s)$, 
the first order of perturbation is 
\[ p^{\prime}=v^2 \rho^{\prime} , \] 
where
\[ v=\sqrt{\Bigl(\frac{\partial p}{\partial \rho} \Bigr)_s} . \]
As a result, we obtain 
\be \label{rhoprime}
\frac{\partial^2 \rho^{\prime}}{\partial t^2}-
v^2 \nabla^D_k \nabla^D_k \rho^{\prime}=0 , \ee
\be  \frac{\partial^2 p^{\prime}}{\partial t^2}-
v^2 \nabla^D_k \nabla^D_k p^{\prime}=0 . \ee
These equations describe the waves in the fractal medium.

\section{Magnetohydrodynamics}

\subsection{Magnetohydrodynamics (MHD) equations}

The hydrodynamic and Maxwell equations for a fractal medium 
\cite{AP2005-2,Plasma2005} are the following. \\
(1) The equation of continuity, 
\be \label{eq1} \Bigl(\frac{d}{dt}\Bigr)_D \rho=-\rho \nabla^D_k u_k . \ee
(2) The equation of balance of density of momentum,  
\be \label{eq2} \rho \Bigl(\frac{d}{dt}\Bigr)_D u_k=
\rho f_k-\nabla^D_k p . \ee
(3) Faraday's law, 
\be \label{eq3}
curl \Bigl(c_1(\gamma,{\bf r}){\bf E} \Bigr)= 
-c_2(d,{\bf r}) \frac{\partial}{\partial t} {\bf B} . \ee
(4) The absence of magnetic monopoles
\be \label{eq4} div \Bigl( c_2(d,{\bf r}) {\bf B} \Bigr)= 0. \ee
(5) Ampere's law, 
\be  \label{eq5} curl \Bigl( c_1(\gamma,{\bf r}){\bf B} \Bigr) = 
\mu_0 c_2(d,{\bf r}) {\bf J} , \ee
where the displacement current is neglected.

Using the Lorenz force density,  
\be
\rho {\bf f}=[{\bf J},{\bf B}] ,
\ee
we get (\ref{eq2}) in the form
\be \rho \Bigl(\frac{d}{dt}\Bigr)_D {\bf u}+ \nabla^D p=
[{\bf J},{\bf B}] . \ee
We assume a linear relationship between the 
${\bf J}$ and ${\bf E}_*$:
\be \label{Om}
{\bf J}({\bf r},t) = \sigma {\bf E}_*({\bf r},t) ,
\ee
where $\sigma$ is the electric conductivity, and
${\bf E}_*$ is an electric field in the moved coordinate system. 
For $|{\bf u}| \ll c$, 
\be \label{E=}
{\bf E}={\bf E}_* - \frac{1}{c}[{\bf u},{\bf B}] .
\ee
From (\ref{Om}), and (\ref{eq5}), we get 
\be \label{E*}
{\bf E}_*=\sigma^{-1} {\bf J}=
\frac{1}{\sigma \mu_0 c_2(d,{\bf r}) }
curl \Bigl( c_1(\gamma,{\bf r}){\bf B} \Bigr) .
\ee
Substitution of (\ref{E=}) into (\ref{eq3}) gives
\[
curl \Bigl(c_1(\gamma,{\bf r})
{\bf E}_* - c_1(\gamma,{\bf r})
\frac{1}{c}[{\bf u},{\bf B}] \Bigr)= 
\]
\be  \label{cur}
=-c_2(d,{\bf r}) \frac{\partial}{\partial t} {\bf B} . \ee
Substituting (\ref{E*}) into (\ref{cur}), we have
\[ curl \Bigl(  
\frac{c_1(\gamma,{\bf r})}{\sigma \mu_0 c_2(d,{\bf r}) }
curl \Bigl( c_1(\gamma,{\bf r}){\bf B} \Bigr)-
\]
\be
- c_1(\gamma,{\bf r}) \frac{1}{c} [{\bf u},{\bf B}] \Bigr)= 
-c_2(d,{\bf r}) \frac{\partial}{\partial t} {\bf B} . \ee
Then, 
\[  c_2(d,{\bf r}) \frac{\partial}{\partial t} {\bf B}=
- curl \left(  
\frac{c_1(\gamma,{\bf r})}{\sigma \mu_0 c_2(d,{\bf r}) }
curl \Bigl( c_1(\gamma,{\bf r}){\bf B} \Bigr) \right)+
\]
\be
+ curl \left( c_1(\gamma,{\bf r})
\frac{1}{c}[{\bf u},{\bf B}] \right) . \ee

As a result, we obtain magnetohydrodynamics (MHD) equations 
for a fractal distribution of charged particles: \\
(1) The equation of continuity, 
\be \label{MHD1} \Bigl(\frac{d}{dt}\Bigr)_D \rho=-\rho \nabla^D {\bf u} . \ee
(2) The equation of balance of density of momentum,  
\be \label{MHD2}
\rho \Bigl(\frac{d}{dt}\Bigr)_D {\bf u}+ \nabla^D p=
[{\bf J},{\bf B}] . \ee
(3) The absence of magnetic monopoles, 
\be \label{MHD3} div \Bigl( c_2(d,{\bf r}) {\bf B} \Bigr)= 0. \ee
(4) The Ampere law, 
\be  \label{MHD4} curl \Bigl( c_1(\gamma,{\bf r}){\bf B} \Bigr) = 
\mu_0 c_2(d,{\bf r}) {\bf J} . \ee
(5) The diffusion equation for the magnetic field, 
\[ 
c_2(d,{\bf r}) \frac{\partial}{\partial t} {\bf B}=
- curl \left(  
\frac{c_1(\gamma,{\bf r})}{\sigma \mu_0 c_2(d,{\bf r}) }
curl \Bigl( c_1(\gamma,{\bf r}){\bf B} \Bigr) \right) + \]
\be \label{MHD5}
+ curl \left( c_1(\gamma,{\bf r})
\frac{1}{c}[{\bf u},{\bf B}] \right) . \ee
We have 11 equations for 11 variables $p$, $\rho$,
${\bf J}$, ${\bf u}$, and ${\bf B}$.

\subsection{Equilibrium from MHD equations}

Let us consider the stationary (equilibrium) states for 
MHD equations. 
The total time derivatives in Eqs. (\ref{MHD2}) and (\ref{MHD4}) 
are equal to zero, and 
\be \label{S1} \nabla^D p= [{\bf J},{\bf B}] , \ee
\be \label{S2} {\bf J}= \frac{1}{\mu_0 c_2(d,{\bf r})} 
curl \Bigl( c_1(\gamma,{\bf r}){\bf B} \Bigr) . \ee
The substitution of (\ref{S2}) into (\ref{S3}) obtains
\be \label{S3} \nabla^D p= \frac{1}{\mu_0 c_2(d,{\bf r})} 
\left[curl \Bigl( c_1(\gamma,{\bf r}){\bf B} \Bigr) ,{\bf B} \right] . \ee
Suppose ${\bf B}=\{0 ,0 ,B_z\}$. Then
\[
curl \Bigl( c_1(\gamma,{\bf r}){\bf B} \Bigr)=
\]
\be
={\bf e}_x \partial_y(c_1(\gamma,{\bf r}) B_z )-
{\bf e}_y \partial_x(c_1(\gamma,{\bf r}) B_z ) ,
\ee
and
\[
\left[curl \Bigl( c_1(\gamma,{\bf r}){\bf B} \Bigr) ,{\bf B} \right]=
\]
\be
=-{\bf e}_x B_z \partial_x(c_1(\gamma,{\bf r}) B_z )-
{\bf e}_y B_z \partial_y(c_1(\gamma,{\bf r}) B_z ) .
\ee
As a result, Eq. (\ref{S1}) gives
\[
\nabla^D_x p=- \frac{1}{\mu_0 c_2(d,{\bf r})}
B_z \partial_x(c_1(\gamma,{\bf r}) B_z ), 
\]
\be
\nabla^D_y p=- \frac{1}{\mu_0 c_2(d,{\bf r})}
B_z \partial_y(c_1(\gamma,{\bf r}) B_z ) . 
\ee
From the definition of $\nabla^D$, we have
\[
\frac{\partial}{\partial x} c_2(d,{\bf r})p
=- \frac{c_3(D,{\bf r})}{\mu_0 c_2(d,{\bf r})}
B_z \frac{\partial}{\partial x} (c_1(\gamma,{\bf r}) B_z ) , 
\]
\be
\frac{\partial}{\partial y} c_2(d,{\bf r})p
=- \frac{c_3(D,{\bf r})}{\mu_0 c_2(d,{\bf r})}
B_z \frac{\partial}{\partial y}(c_1(\gamma,{\bf r}) B_z ). 
\ee
Using $A\partial B=\partial(AB)-B\partial A$, we get
\[
\frac{\partial}{\partial x} \left( c_2(d,{\bf r})p+
\frac{c_3(D,{\bf r}) c_1(\gamma,{\bf r})}{\mu_0 c_2(d,{\bf r})} B^2_z
\right)= 
\]
\[
=c_1(\gamma,{\bf r}) B_z \frac{\partial}{\partial x}
\left( \frac{c_3(D,{\bf r})}{\mu_0 c_2(d,{\bf r})} B_z \right),
\]
\[
\frac{\partial}{\partial y} \left( c_2(d,{\bf r})p+
\frac{c_3(D,{\bf r}) c_1(\gamma,{\bf r})}{\mu_0 c_2(d,{\bf r})} B^2_z
\right)= 
\]
\be
=c_1(\gamma,{\bf r}) B_z \frac{\partial}{\partial y}
\left( \frac{c_3(D,{\bf r})}{\mu_0 c_2(d,{\bf r})} B_z \right), 
\ee
\be
\frac{\partial}{\partial y} \left( c_2(d,{\bf r})p \right)=0. 
\ee
As a result, we obtain 
\be
c_2(d,{\bf r})p+
\frac{c_3(D,{\bf r}) 
c_1(\gamma,{\bf r})}{\mu_0 c_2(d,{\bf r})} B^2_z=const .
\ee
This equilibrium equation exists only if 
\be
B_z \sim \frac{\mu_0 c_2(d,{\bf r})}{c_3(D,{\bf r})} . \ee

It is easy to see that we do not have the usual invariants 
for the fractal distribution of charged particles. 
Therefore equilibrium on the fractal exists 
for the magnetic field that satisfies the power law relation
\be \label{Bz}
B_z \sim R^{d-D+1} .
\ee 
For the distribution with an integer Hausdorff dimension, 
we have the usual relation \cite{Kad}.
The typical turbulent media could be of fractal structure, 
and the corresponding equations should be changed to include
the fractal features of the media.
Therefore, the equilibrium of the fractal turbulent medium exists 
for the magnetic field with the power law relation (\ref{Bz}).

\section{Conclusion}

Typical turbulent media could be of a fractal structure, 
and the corresponding equations should be changed to include
the fractal features of the media.
Magnetohydrodynamics equations for 
the fractal distribution of charged particles are suggested.
The fractional integrals are used to describe fractal distribution.
These integrals are considered as approximations of
integrals on fractals. 
Using the fractional generalization of the integral Maxwell equation
and the integral balance equations, we derive
the magnetohydrodynamics equations.
Equilibrium states for these equations are discussed. 
The equilibrium for fractal turbulent media can exists 
if the magnetic field satisfies the power law relation.


\section{Appendix: Fractional Gauss's theorem}

Let us derive the fractional 
generalization of Gauss's theorem, 
\be \label{AAA} 
\int_{\partial W} ({\bf J}({\bf r},t), d{\bf S}_2) 
=\int_W div( {\bf J}({\bf r},t) ) dV_3 , 
\ee
where the vector ${\bf J}({\bf r},t)=J_k{\bf e}_k$ is a field, 
and $div( {\bf J})={\partial {\bf J}}/{\partial {\bf r}}= 
{\partial J_k}/{\partial x_k}$.
Here, we mean the sum on the repeated index
$k$ from 1 to 3. Using 
\[ d{\bf S}_d=c_2 (d,{\bf r})d{\bf S}_2 , \quad 
c_2(d,{\bf r})= \frac{2^{2-d}}{\Gamma(d/2)} |{\bf r}|^{d-2} , \]
we get
\[ \int_{\partial W} ({\bf J}({\bf r},t),d{\bf S}_d) 
=\int_{\partial W}  c_2(d,{\bf r})  ({\bf J}({\bf r},t) , d{\bf S}_2) . \]
Note that $c_2(2,{\bf r})=1$ for $d=2$. 
Using (\ref{AAA}), we get 
\[ \int_{\partial W}  c_2(d,{\bf r}) ({\bf J}({\bf r},t), d{\bf S}_2) =
\int_W  div(c_2(d,{\bf r}) {\bf J}({\bf r},t)) dV_3 . \]
The relation
\[ dV_D=c_3 (D,{\bf r})dV_3 , \quad 
c_3(D,{\bf r})= \frac{2^{3-D} \Gamma(3/2)}{\Gamma(D/2)} |{\bf r}|^{D-3}  \]
in the form $dV_3=c^{-1}_3(D,{\bf r}) dV_D$
allows us to derive the fractional generalization of Gauss's theorem:
\[ \int_{\partial W} ({\bf J}({\bf r},t), d{\bf S}_d)=
\]
\[
=\int_W c^{-1}_3(D,{\bf r}) 
div \Bigr( c_2(d,{\bf r}) {\bf J}({\bf r},t) \Bigr) \ dV_D .\]
Analogously, we can get the fractional generalization
of Stokes's theorem in the form
\[ \oint_L ({\bf E},d{\bf l}_{\gamma})=
\int_S  c^{-1}_2(d,{\bf r})
(curl(c_1(\gamma,{\bf r}){\bf E}), d{\bf S}_d) , \]
where 
\[ c_1(\gamma,{\bf r})=
\frac{2^{1-\gamma}\Gamma(1/2)}{\Gamma(\gamma/2)}|{\bf r}|^{\gamma-1} . \]


\end{document}